\documentclass[preprint,showpacs,preprintnumbers,amsmath,amssymb]{revtex4}
\usepackage{graphicx}% Include figure files
\usepackage{dcolumn}% Align table columns on decimal point
\usepackage{bm}% bold math

\begin{document}

\title{Strong interference effects in the resonant Auger decay of atoms induced by intense X-Ray fields}

\author{\firstname{Philipp~V.} \surname{Demekhin}}
\email{philipp.demekhin@pci.uni-heidelberg.de}
\altaffiliation[\\On leave from: ]{Rostov State University of TC,
Narodnogo Opolche\-niya square 2, 344038, Rostov-on-Don, Russia}
\affiliation{Theoretische Chemie, Physikalisch-Chemisches
Institut, Universit\"{a}t  Heidelberg, Im Neuenheimer Feld 229,
D-69120 Heidelberg, Germany}

\author{\firstname{Lorenz~S.} \surname{Cederbaum}}
\affiliation{Theoretische Chemie, Physikalisch-Chemisches Institut, Universit\"{a}t Heidelberg,
Im Neuenheimer Feld 229, D-69120 Heidelberg, Germany}

\date{\today}

\begin{abstract}
The theory of resonant Auger decay of atoms in a high intensity coherent X-ray pulse is presented. The theory includes the coupling between the ground state and the resonance due to an intense X-ray pulse, taking into account the decay of the resonance and the direct photoionization of the ground state, both populating the final ionic states coherently. The theory also considers the impact of the direct photoionization of the resonance state itself which typically populates highly-excited ionic states. The combined action of the resonant decay and of the direct ionization of the ground state in the field induces a non-hermitian time-dependent coupling between the ground and the `dressed' resonance stats. The impact of these competing processes on the total electron yield and on the 2s$^2$2p$^{4}(^1\mathrm{D})$3p~$^2$P spectator and  2s$^1$2p$^{6}$~$^2$S participator Auger decay spectra of the Ne  1s$\to$3p resonance is investigated. The role of the direct photoionization of the ground state and of the resonance increases dramatically with the field intensity. This results in strong interference effects with distinct patterns in the electron spectra, different for the participator and spectator final states.
\end{abstract}

\pacs{33.20.Xx, 32.80.Hd, 41.60.Cr, 82.50.Kx}
%\keywords{Spectra induced by strong-field or attosecond laser irradiation, Auger effect and inner-shell excitation or ionization, Free-electron lasers (see also 52.59.Rz Free-electron devices擁n plasma physics),Processes caused by x-rays or y-rays }

\maketitle

\section{Introduction}
\label{sec:intro}

The resonant Auger (RA) effect of atoms, discovered more than 30 years ago \cite{Eberhardt78,Brown80}, has been extensively studied over the past decades experimentally by utilizing conventional synchrotron radiation sources and theoretically (see, e.g., the review \cite{Armen00} and references therein). In contrast to X-ray photoelectron spectroscopy (XPS), the RA decay spectra are recorded at the exciting-photon energy tuned to a resonance. In the vicinity of core-excitations of atoms, the resonant photoionization channel (i.e., the excitation and Auger decay of a highly-excited electronic state) dominates over the direct channel (i.e., non-resonant photoionization) populating the same final ionic state, owing to the large energy of the exciting X-ray photons. As a result, the RA decay electron spectra provide uniquely important information on the electronic structure and decay mechanisms of highly-excited electronic states.

The impact of the direct photoionization on the RA spectra is nevertheless not negligible. There are several theoretical and experimental studies indicating the relevance of the weak direct photoionization channel for the interpretation of RA spectra of atoms and molecules \cite{Camilloni96,Kukk97,Saito00,Lagutin03KrAugLet,Carravetta97,Feifel06,Kukk99,LagPRL03,Lagutin03Rabig,Demekhin09CstarO,Demekhin09COstar,Demekhin10prlNO,Demekhin10jpbNOlarge}. Perceptible fingerprints of the interference between the dominant resonant and the weak direct photoionization channels were observed in the angular-averaged RA decay spectra and in the branching ratios of different decay channels scanned across the resonance \cite{Camilloni96,Kukk97,Saito00,Lagutin03KrAugLet,Carravetta97,Feifel06}. This interference manifests itself even more distinctly in the angular-resolved RA decay spectra  \cite{Kukk99,LagPRL03,Lagutin03Rabig,Demekhin09CstarO,Demekhin09COstar,Demekhin10prlNO,Demekhin10jpbNOlarge}. As demonstrated in these works, the weak direct photoionization channel leads to a broad photon-energy dependent dispersion of the RA electron angular distribution and residual ion polarization parameters.

The advent of X-ray Free Electron Lasers (XFELs)  \cite{XFEL1,XFEL2} raises the fundamental question of how the well-studied processes of interaction of matter with electromagnetic field will be modified under extreme field conditions. Compared to conventional synchrotron radiation sources, the intensities of the electromagnetic fields generated by XFELs are by several orders of magnitude stronger, and the durations of the light pulses can be made to be similar to the typical lifetimes of highly-excited electronic states (typically a few femtoseconds). The first theoretical studies of RA decay of atoms exposed to strong X-ray pulses \cite{Rohringer08,Liu10,Sun10} showed that the stimulated emission from the resonance back to the ground state starts to compete with the Auger decay. The interplay between the resonant excitation and stimulated emission results in Rabi oscillations between the ground state and the resonance within its Auger decay lifetime and pulse duration \cite{Rohringer08,Liu10}, which leads to spectacular modifications of the RA spectra. It has been demonstrated in Refs.~\cite{Liu10,Sun10} that in the presence of such strong electromagnetic fields the direct photoionization of the ground state may start to play an important role.

Obviously, a final ionic state is populated coherently by both the direct photoionization and the resonant Auger decay thus naturally inducing interference effects in the electron spectra. On the other hand, as discussed in \cite{Liu10,Sun10}, there will be leakage of the population of the ground state by the direct photoionization into all possible final ionic states (total photoionization). We shall demonstrate here that during the duration of a strong pulse the direct photoionization of the resonant state itself to produce preferably highly-excited ionic states may also take place and compete with its decay via Auger and stimulated emission to the ground state. It has not yet been shown and understood how all these processes evoked by a strong field influence the total electron yield and the spectra and filling this gap is the major goal of this paper.

The derivation of the theory of electron spectra of atoms exposed to strong X-ray pulses is presented in Secs.~\ref{sec:thyR}--\ref{sec:thySTAR}. The readers not interested in the details of the derivation are referred to Sec.~\ref{sec:thyFIN}, where the final set of equations is collected and analyzed. In Sec.~\ref{sec:res}, the derived theory is applied to the Auger decay of the 1s$^1$2s$^2$2p$^6$3p$^1$~$^1$P  resonance of the Ne atom into the conceptually different spectator and participator decay channels. We conclude with a brief summary.

\section{Theory}
\label{sec:thy}

The currently operating XFEL, Linac Coherent Light Source (LCLS), does not produce so far a monochromatic radiation and its X-ray pulses consist of many spikes with random fluctuations of the frequency, phase and amplitude \cite{XFEL1}. The impact of these problems on the RA effect has been studied in  \cite{Rohringer08}. In addition, the X-ray pulse properties are compressed through the propagation in a resonant medium \cite{Sun10}.

\begin{figure}
\includegraphics[scale=0.6]{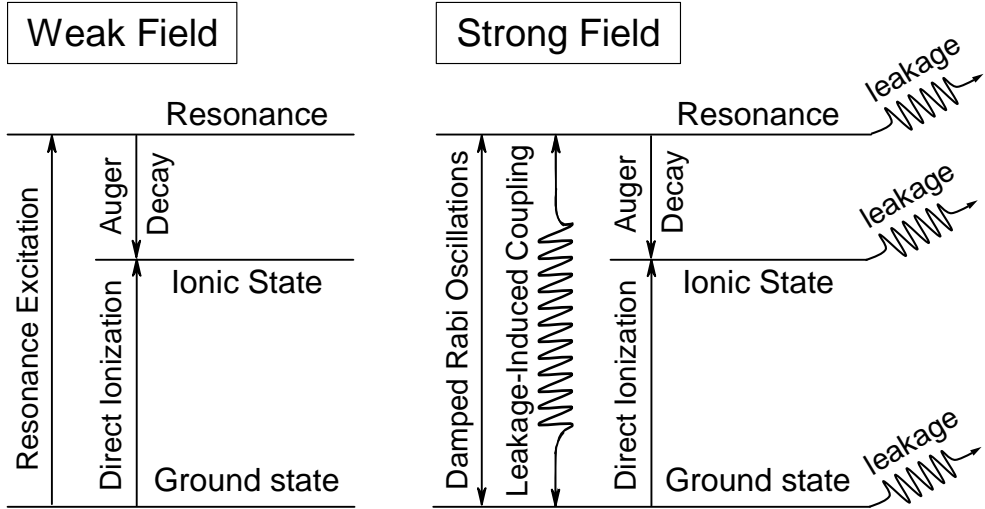}
\caption{Schematic representation of the resonant Auger effect of an atom in weak (left panel) and strong (right panel) fields. \emph{Left panel}: In a weak electromagnetic field, the resonance is excited and decays populating a final ionic state which is also populated by the direct photoionization of the ground state. Both pathways coherently superimpose, but due to the high X-ray frequency, the resonant pathway dominates by far. After the pulse expires, the residual population of the ground state remains nearly 1. \emph{Right panel}:  In the case of a strong field, the ground state and the resonance are strongly coupled by the field. The usual direct coupling which is known to induce Rabi oscillations between these two states is modified by an interesting term which is indicated by a vertical helix double-arrow.  The ground state, the resonance and also the produced ionic states are subject to time-dependent losses (leakages) due to photoionization as indicated by the inclined helix arrow from the respective state. The modified coupling and the leakages influence the RA effect and lead to a damping of the  Rabi oscillations. After the pulse expires, the population of the ground state can be very small.}
\label{fig:scheme}
\end{figure}

In the present work we concentrate on the physics a single atom undergoes when exposed to a coherent and monochromatic X-ray pulse. The knowledge of this physics is prerequisite for further studies. For simplicity we assume a pulse linearly polarized along the $z$ axis with the field:
\begin{equation}
\label{eq:e_vector}
\mathcal{E}(t)=\mathcal{E}_0(t)\cos\omega t=\mathcal{E}_0 \,g(t) \cos\omega t.
\end{equation}
Here $\mathcal{E}_0$ is the peak amplitude, and the pulse-shape function $g(t)$ varies slowly on the timescale of $2\pi/\omega$. The cycle-averaged intensity of the field is given in atomic units via (1 a.u. = 6.43641$\times 10^{15}$ W/cm$^2$)
\begin{equation}
\label{eq:intens}
I(t)=\frac{1}{8\pi\alpha}\left\{\mathcal{E}_0 \,g(t)\right\}^2 ,
\end{equation}
where $ \alpha=1/137.036$ is the fine structure constant.

The processes relevant for the present study are schematically shown in  Fig.~\ref{fig:scheme}. In a weak electromagnetic field (left picture), the resonance is excited and decays populating a final ionic state which is also populated by the direct photoionization of the ground state. Both pathways coherently superimpose, but due to the high X-ray frequency, the resonant pathway dominates by far. As the pulse expires, the non-resonant pathway expires as well, and the population of the ground state remains close to the unity. In the case of a strong field (right picture), the ground state and the resonance are strongly coupled by the field. We shall demonstrate below that the usual direct coupling which is known to induce Rabi oscillations between these two states is modified by an interesting term derived in Sec.~\ref{sec:thyRD} and discussed in Sec.~\ref{sec:thyFIN}. The modification of the coupling between the states is indicated in Fig.~\ref{fig:scheme} by a vertical helix double-arrow.  In addition, there are two mechanisms of losses relevant to the RA effect in strong fields: The total photoionization of the ground state and of the resonance lead to a leakages of the population of the respective states, as indicated in Fig.~\ref{fig:scheme} by the inclined helix arrows from the ground state and from the resonance, respectively.  These two loss mechanisms have interesting impact on the RA effect. The leakages and the modified coupling  all lead to damped Rabi oscillations of the population between the ground state and the resonance. Of course, these processes, except of the Auger decay itself, are operative only when the pulse is on. After the pulse expires, the final population of the ground state can be much smaller than 1.

In the presence of a strong pulse, multiple ionization of the atom takes place as well  \cite{XFEL2}. There are several mechanisms responsible for that. One is the ionization of the final ionic states populated via the RA effect (indicated by the inclined helix arrow from the ionic state in Fig.~\ref{fig:scheme}). The ionization of the resonance mentioned above is another. This ionization produces preferentially highly-excited ionic states, most of them with a core electron missing as in the resonance state. These ionic states are then likely to undergo Auger decay producing thereby doubly-ionized or even higher ionized atoms. All the processes discussed here can, in principle, be experimentally separated from each other by measuring the energies of the ejected electrons.

Below we concentrate on the processes shown in the scheme of Fig.~\ref{fig:scheme}. We mention that the leakage from the ionic states  is not relevant for the RA effect as long as one measures electrons and does not intend to measure the ions. To describe the RA effect theoretically as a function of time, one needs to solve the time-dependent Schr\"{o}dinger equation for the atom and its interaction with the field  (atomic unis $e=m_e=\hbar=1$ are used throughout)
\begin{equation}
\label{eq:hamilt}
i\dot{\Psi}(t)=\hat{H}(t)\Psi(t)=\left(\hat{H}_0+ \hat{z}\, \mathcal{E}(t)\right)\Psi(t) . 
\end{equation} 
The present approach to solve Eq.~(\ref{eq:hamilt})  is similar to that reported in \cite{Lytitia10}. For transparency of  presentation we derive below step by step the equations for the different mechanisms depicted in Fig.~\ref{fig:scheme} and collect all results and discuss them in Sec.~\ref{sec:thyFIN}.

\subsection{Resonant channel}
\label{sec:thyR}

As a first approximation, we include the resonant  channel, consisting of the excitation and Auger decay of the resonance  and the stimulated emission from the resonance. We start with the following ansatz for the total wave function of the  three-energy-level system: the ground state $\vert I \rangle$, the resonance $\vert R \rangle$, and the final ionic state plus   Auger electron  $\vert F \varepsilon \rangle$, with the energies $E_I$, $E_R$, $E_F$ and $\varepsilon$, respectively 
\begin{equation}
\label{eq:anzatzR}
\Psi(t)= a_I(t)\vert I \rangle+\widetilde{a}_R(t)\vert R \rangle+\int \widetilde{a}_F (t,\varepsilon) \vert F \varepsilon \rangle d\varepsilon.
\end{equation}
Here $a_I(t)$, $\widetilde{a}_R(t)$, and $\widetilde{a}_F (t,\varepsilon)$ are the time-dependent amplitudes for the population of the  $\vert I \rangle$, $\vert R \rangle$, and  $\vert F \varepsilon \rangle$ levels, respectively. The transitions between the atomic states are described by the following matrix elements of the total Hamiltonian~(\ref{eq:hamilt})
\begin{subequations}
\label{eq:couplingsR}
\begin{equation}
\label{eq:couplingsRexc}
\langle R\vert \hat{H}(t) \vert I\rangle= \langle R\vert \hat{z}\vert I\rangle \frac{\mathcal{E}_0 \,g(t)}{2}\, e^{-i\omega t}= D_R(t)\, e^{-i\omega t},
\end{equation}
\begin{equation}
\label{eq:couplingsRAD}
\langle F\varepsilon \vert \hat{H}(t) \vert R\rangle=  V.
\end{equation}
\end{subequations}
In   Eq.~(\ref{eq:couplingsRexc}), the rotating wave approximation \cite{Griffiths94} has already been utilized, and, in contrast to the rapidly oscillating factor $e^{-i\omega t}$, the function  $D_R(t)$ varies slowly on the timescale of $2\pi/\omega$. Here and below, all transition matrix elements are assumed to  vary slowly with the energy across the resonance  and are replaced by their mean values.

By substituting ansatz (\ref{eq:anzatzR}) in the  time-dependent Schr\"{o}dinger equation and projecting the result onto each state, one obtains the following set of differential equations for the time-dependent amplitudes 
\begin{subequations}
\label{eq:CDE_R}
\begin{equation}
\label{eq:CDE_R1}
i\dot{a}_I(t)=E_Ia_I(t)+D^\dag_R(t)\, e^{+i\omega t}\widetilde{a}_R(t)
\end{equation}
\begin{equation}
\label{eq:CDE_R2}
i\dot{\widetilde{a}}_R(t)=D_R(t)\, e^{-i\omega t}{a}_I(t)+E_R\,\widetilde{a}_R(t) +\int  V^\dag \,\widetilde{a}_F (t,\varepsilon) d\varepsilon
\end{equation}
\begin{equation}
\label{eq:CDE_R3}
i\dot{\widetilde{a}}_F (t,\varepsilon)= V\widetilde{a}_R(t)+ \left(E_F+\varepsilon\right)\widetilde{a}_F (t,\varepsilon).
\end{equation}
\end{subequations}
It is possible to eliminate the rapidly oscillating factors $e^{\pm i\omega t}$  from the system of equations~(\ref{eq:CDE_R}). For this purpose we redefine the time-dependent amplitudes as follows
\begin{equation}
\label{eq:redefineR}
{a}_R(t)= \widetilde{a}_R(t)\,e^{+i\omega t}~~~ \mathrm{and}~~~ {a}_F (t,\varepsilon)=\widetilde{a}_F (t,\varepsilon) \,e^{+i\omega t}.
\end{equation}
In the local approximation \cite{Cederbaum81,Domcke91}, the integral in the right part of Eq.~(\ref{eq:CDE_R2}) is equal to
\begin{equation}
\label{eq:part_width}
\int  V^\dag \,{a}_F (t,\varepsilon)d\varepsilon=-i\pi \vert V \vert^2 {a}_R(t),
\end{equation}
as has been explicitly demonstrated  in Ref.~\cite{Pahl99ZPD} and will also be outlined in the next subsection for the case of the direct ionization channel.   Finally, when several Auger decay channels of the resonance into different final ionic states $\vert F_j \varepsilon_j \rangle$ exist, Eq.~(\ref{eq:part_width}) must be modified as follows
\begin{equation}
\label{eq:tot_width}
\sum _j\int  V^\dag_{j} {a}_{F_j} (t,\varepsilon_j)d\varepsilon_j=-i\pi\sum _j \vert V_{j} \vert^2{a}_R(t)=-\frac{i}{2}\Gamma_A{a}_R(t),
\end{equation}
where $\Gamma_A$ is the total rate for the Auger decay of the resonance.

In the above designations, the set of coupled differential equations for the time-dependent amplitudes reads
\begin{subequations}
\label{eq:CDE_RN}
\begin{equation}
\label{eq:CDE_RN1}
i\dot{a}_I(t)=E_Ia_I(t)+D^\dag_R(t) \,{a}_R(t)
\end{equation}
\begin{equation}
\label{eq:CDE_RN2}
i\dot{ {a}}_R(t)=D_R(t)\,  {a}_I(t)+(E_R -\frac{i}{2}\Gamma_A-\omega ){a}_R(t) 
\end{equation}
\begin{equation}
\label{eq:CDE_RN3}
i\dot{a}_{F_j} (t,\varepsilon_j)= V_{j}\,{a}_R(t)+ \left(E_{F_j}+\varepsilon _j-\omega\right)a_{F_j} (t,\varepsilon_j).
\end{equation}
\end{subequations}
The system of Eqs.~(\ref{eq:CDE_RN}) is   consistent with the equations reported in \cite{Rohringer08}. It includes the coupling of the ground electronic state of energy $E_I$ with the `dressed' resonance state of   energy $E_R-\omega$ by the matrix element $D_R(t)$, as well as the leakage of the resonance population with   total rate  $\Gamma_A$ into the   `dressed' continuum final states with the  energies $E_{F_j}+\varepsilon_j-\omega$ via the partial Coulomb matrix elements $V_{j}$.  The leakage of the resonant population caused by the Auger decay mechanism is referred to hereafter  as the `\emph{RA-leakage}'.

\subsection{Direct channel}
\label{sec:thyD}

In this subsection, only the  direct population of the final ionic state  $\vert F \varepsilon \rangle$ from  the ground state $\vert I \rangle$ via the   transition matrix element
\begin{equation}
\label{eq:directME}
\langle F\varepsilon \vert \hat{H}(t) \vert I\rangle= \langle F\varepsilon \vert \hat{z}\vert I\rangle \frac{\mathcal{E}_0 \,g(t)}{2}\, e^{-i\omega t}= D_D(t)\, e^{-i\omega t}
\end{equation}
is investigated.  We apply the following ansatz for the total wave function of the  two-level system:
\begin{equation}
\label{eq:anzatzD}
\Psi(t)= a_I(t)\vert I \rangle+ \int \widetilde{a}_F (t,\varepsilon) \vert F \varepsilon \rangle d\varepsilon.
\end{equation}
After redefining the $\widetilde{a}_F (t,\varepsilon)$  via Eq.~(\ref{eq:redefineR}), the set of differential equation for the time-dependent amplitudes  reads
\begin{subequations}
\label{eq:CDE_D}
\begin{equation}
\label{eq:CDE_D1}
i\dot{a}_I(t)=E_Ia_I(t)+  \int  D^\dag_D(t)\,  {a}_F (t,\varepsilon)d\varepsilon
\end{equation}
\begin{equation}
\label{eq:CDE_D2}
i\dot{ {a}}_F (t,\varepsilon)=  D_D(t)\,a_I(t) + \left(E_F+\varepsilon-\omega\right) {a}_F (t,\varepsilon).
\end{equation}
\end{subequations}

The formal solution of Eq.~(\ref{eq:CDE_D2}) is given by
\begin{equation}
\label{eq:solut_D1}
{a}_F (t,\varepsilon)=-i \int_{-\infty}^t D_D(t^\prime) a_I(t^\prime)  \,  e^{-i(E_F+\varepsilon-\omega )(t-t^\prime)} dt^\prime.
\end{equation}
With the help of this expression, the integral   on the right-hand side of Eq.~(\ref{eq:CDE_D1}) can be rewritten as 
\begin{equation}
\label{eq:solut_D2}
 \int  D^\dag_D(t)\,  {a}_F (t,\varepsilon)d\varepsilon =-i \int_{-\infty}^t  K(t-t^\prime) a_I(t^\prime)  dt^\prime,
\end{equation}
with the non-local kernel
\begin{equation}
\label{eq:solut_D3}
 K(t-t^\prime)= \int  D^\dag_D(t)  \, D_D(t^\prime)   \,  e^{-i(E_F+\varepsilon-\omega )(t-t^\prime)}   d\varepsilon .
\end{equation}
The kernel~(\ref{eq:solut_D3}) accounts for the effect of coupling   the ground electronic state with the  continuum of final states. In the local approximation \cite{Cederbaum81,Domcke91}, the integrations over the energy and time in Eqs.~(\ref{eq:solut_D3}) and (\ref{eq:solut_D2}) can be performed analytically  (see Appendix \ref{sec:app} for details). The final result reads 
\begin{equation}
\label{eq:solut_D4}
 \int  D^\dag_D(t)\,  {a}_F (t,\varepsilon)d\varepsilon =-i \pi \vert D_D(t)\vert^2  a_I(t) .
\end{equation}
In the presence of several  final ionic states $\vert F_j \varepsilon_j \rangle$, Eq.~(\ref{eq:solut_D4}) must be modified as follows
\begin{equation}
\label{eq:solut_D5}
\sum _j  \int  D^\dag_{D_j}(t)\,{a}_{F_j} (t,\varepsilon_j)d\varepsilon_j=-i\pi\sum _j  \vert D_{D_j}(t)\vert^2  a_I(t) = -\frac{i}{2}\Gamma_{ph}(t){a}_I(t),
\end{equation}
where $\Gamma_{ph}(t)$ is the time-dependent total probability for the direct photoionization of the ground state. This $\Gamma_{ph}(t)$ is identical with the $\gamma_{ph}(t)$ introduced in Ref.~\cite{Liu10} (see discussion around Eq.~(16) in this reference).

In the above designations, the set of equations for the time-dependent amplitudes reads
\begin{subequations}
\label{eq:CDE_DN}
\begin{equation}
\label{eq:CDE_DN1}
i\dot{a}_I(t)=(E_I-\frac{i}{2}\Gamma_{ph}(t))\,a_I(t) 
\end{equation}
\begin{equation}
\label{eq:CDE_DN2}
i\dot{ {a}}_{F_j} (t,\varepsilon_j)=  D_{D_j}(t)\,a_I(t) + \left(E_{F_j}+\varepsilon_j-\omega\right) {a}_{F_j} (t,\varepsilon_j).
\end{equation}
\end{subequations}
The system of Eqs.~(\ref{eq:CDE_DN})   includes the leakage of the population of the ground state with the total probability for the direct photoionization $\Gamma_{ph}(t)$, as well as  the direct population of the `dressed' continuum final states with the energy $E_{F_j}+\varepsilon_j-\omega$ via the partial matrix elements $ D_{D_j}(t)$. The leakage of the population of the ground state owing to the non-resonant direct photoionization channel is referred to hereafter as the `\emph{GS-leakage}'. The GS-leakage  is  time-dependent, taking place only during the pulse duration (see Eq.~(\ref{eq:directME}) for the definition of the function $D_D(t)$).

\subsection{Interference of resonant and direct channels}
\label{sec:thyRD}

In order to account for the interference between the resonant and direct ionization channels, we apply the ansatz~(\ref{eq:anzatzR}) for the total wave function of the three-level system with the coupling matrix elements given by (\ref{eq:couplingsR}) and (\ref{eq:directME}). The set of equation for the time-dependent amplitudes can be obtained as described above. After the redefinitions (\ref{eq:redefineR}) it reads
\begin{subequations}
\label{eq:CDE_RD}
\begin{equation}
\label{eq:CDE_RD1}
i\dot{a}_I(t)=E_Ia_I(t)+D^\dag_R(t)\,  {a}_R(t)+\int  D^\dag_D(t)\,  {a}_F (t,\varepsilon)d\varepsilon
\end{equation}
\begin{equation}
\label{eq:CDE_RD2}
i\dot{{a}}_R(t)=D_R(t)\,  {a}_I(t)+(E_R-\omega) \,{a}_R(t) +\int  V^\dag \,{a}_F (t,\varepsilon) d\varepsilon
\end{equation}
\begin{equation}
\label{eq:CDE_RD3}
i\dot{{a}}_F (t,\varepsilon)= D_D(t)\,a_I(t)+ V{a}_R(t) + \left(E_F+\varepsilon-\omega\right){a}_F (t,\varepsilon).
\end{equation}
\end{subequations}
The formal solution of Eq.~(\ref{eq:CDE_RD3}) is given by
\begin{equation}
\label{eq:solut_DR1}
{a}_F (t,\varepsilon)=-i \int_{-\infty}^t \left\{ D_D(t^\prime)\, a_I(t^\prime)+V {a}_R(t^\prime) \right\} 
   e^{-i(E_F+\varepsilon-\omega )(t-t^\prime)} dt^\prime.
\end{equation}
Utilizing this expression and implying the local approximation  the integrations on the right-hand sides of Eqs.~(\ref{eq:CDE_RD1}) and (\ref{eq:CDE_RD2}) can be performed analytically, as shown in Appendix~\ref{sec:app} for the case of the direct channel.  The final results read
\begin{equation}
\label{eq:solut_DR2}
 \int  D^\dag_D(t)\,  {a}_F (t,\varepsilon)d\varepsilon =-i \pi \left\{\vert D_D(t)\vert^2   a_I(t) + D^\dag_D(t)V{a}_R(t) \right\},
\end{equation}
\begin{equation}
\label{eq:solut_DR3}
\int V^\dag \,{a}_F (t,\varepsilon) d\varepsilon=-i\pi\left\{ D_D(t)V^\dag {a}_I(t)+  \vert V \vert^2 {a}_R(t)\right\}.
\end{equation}

In the presence of several  final ionic states $\vert F_j \varepsilon_j \rangle$, and in the above designations, the set of  Eqs.~(\ref{eq:CDE_RD})  takes the following final form
\begin{subequations}
\label{eq:CDE_RDN}
\begin{equation}
\label{eq:CDE_RDN1}
i\dot{a}_I(t)=\left(E_I-\frac{i}{2}\Gamma_{ph}(t)\right)\,a_I(t) 
 +\left(D^\dag_R(t)-\frac{i}{2} W^\dag(t)\right)\,{a}_R(t)
\end{equation}
\begin{equation}
\label{eq:CDE_RDN2}
i\dot{ {a}}_R(t)=\left(D_R(t)-\frac{i}{2} W(t)\right)\, {a}_I(t) +\left(E_R -\frac{i}{2}\Gamma_A-\omega \right){a}_R(t) 
\end{equation}
\begin{equation}
\label{eq:CDE_RDN3}
i\dot{a}_{F_j} (t,\varepsilon_j)=D_{D_j}(t)\,a_I(t) + V_{j}\,{a}_R(t) + \left(E_{F_j}+\varepsilon _j-\omega\right)a_{F_j} (t,\varepsilon_j),
\end{equation}
\end{subequations}
with the function $W(t)$  defined via
\begin{equation}
\label{eq:solut_DR5}
W(t)=2\pi \sum_j D_{D_j}(t)V^\dag_{j}.
\end{equation}
In addition to the GS-leakage in Eq.~(\ref{eq:CDE_RDN1}) and   the RA-leakage in  Eq.~(\ref{eq:CDE_RDN2}), the system of Eqs.~(\ref{eq:CDE_RDN}) incorporates  the  populations of the final ionic continuum state via the resonant and direct ionization channels coherently (the sum of  two corresponding  amplitudes on the right-hand side of Eq.~(\ref{eq:CDE_RDN3})). In the presence of the direct  ionization of the ground state, the coupling matrix element  between the  ground and `dressed' resonance states is modified by the time-dependent function (\ref{eq:solut_DR5})  (cf Eqs.~(\ref{eq:CDE_RN}) and (\ref{eq:CDE_RDN})). In additional to the coupling caused by the field  $D_R(t)$  there is now the non-hermitian term $-\frac{i}{2} W(t)$, induced by the GS-leakage mechanism.  This term is referred below as the Leakage-Induced Complex coupling  `\emph{LIC-coupling}'.  We note that the whole coupling matrix element is now  non-hermitian.

\subsection{Direct ionization of resonance}
\label{sec:thySTAR}

It is very unlikely that the energy tuned for the resonant core excitation will match the one required for  further resonant excitation of the resonance. However, the energy of radiation is above the  ionization threshold  of the outer shells of the resonantly excited state. This may lead to a non-resonant ionization of the resonance state by preferably populating   highly-excited ionic states (see Fig.~\ref{fig:scheme}). In order to incorporate this mechanism into the theory, we extend the ansatz (\ref{eq:anzatzR}) by the following term 
\begin{equation}
\label{eq:anzatzRRR1}
\int \widetilde{a}_{F_\ast} (t,\varepsilon_\ast) \vert F_\ast \varepsilon_\ast \rangle d\varepsilon_\ast
\end{equation}
and introduce the transition matrix element between the resonant state and the excited  state of the ion plus electron $\vert F_\ast \varepsilon_\ast \rangle$ with   energies $E_{F_\ast}$ and $\varepsilon_\ast$
\begin{equation}
\label{eq:couplingsRRR2}
\langle   F_\ast \varepsilon_\ast  \vert \hat{H}(t) \vert R\rangle= \langle F_\ast \varepsilon_\ast  \vert \hat{z}\vert R\rangle \frac{\mathcal{E}_0 \,g(t)}{2}\, e^{-i\omega t}= D_\ast(t)\, e^{-i\omega t}.
\end{equation}
Similar to Eq.~(\ref{eq:redefineR}), the time-dependent amplitude $\widetilde{a}_{F_\ast}(t,\varepsilon_\ast)$ must be redefined as
\begin{equation}
\label{eq:couplingsRRR3}
{a}_{F_\ast}(t,\varepsilon_\ast)=\widetilde{a}_{F_\ast}(t,\varepsilon_\ast) \,e^{+2i\omega t}.
\end{equation}

Incorporation of the non-resonant ionization of the resonance results in the modification  of Eq.~(\ref{eq:CDE_RDN2}). In the case of many direct ionization channels of the resonance into  $\vert F_{\ast_j} \varepsilon_{\ast_j} \rangle$ states, Eq.~(\ref{eq:CDE_RDN2}) will contain the following additional term on the right-hand side
\begin{equation}
\label{eq:couplingsRRR4}
\sum _j  \int  D^\dag_{\ast_j}(t)\,{a}_{F_{\ast_j}} (t,\varepsilon_{\ast_j})d\varepsilon_{\ast_j}.
\end{equation}
The set of Eqs.~(\ref{eq:CDE_RDN}) must also be extended by the following equation for the amplitude ${a}_{F_{\ast_j}}(t,\varepsilon_{\ast_j})$
\begin{equation}
\label{eq:couplingsRRR5}
i\dot{a}_{F_{\ast_j}}(t,\varepsilon_{\ast_j})=  D_{\ast_j}(t)\,a_R(t)  + \left(E_{F_{\ast_j}}+\varepsilon_{\ast_j}-2\omega\right) {a}_{F_{\ast_j}} (t,\varepsilon_{\ast_j}).
\end{equation}
One can decouple Eq.~(\ref{eq:couplingsRRR5}) from the whole set of equations. By substituting the formal solution of Eq.~(\ref{eq:couplingsRRR5}) into the integral~(\ref{eq:couplingsRRR4}) and implying the local approximation, one obtains    
\begin{equation}
\label{eq:couplingsRRR6}
\sum _j  \int  D^\dag_{\ast_j}(t)\,{a}_{F_{\ast_j}} (t,\varepsilon_{\ast_j})d\varepsilon_{\ast_j}=-i\pi\sum _j \vert D_{\ast_j}(t)\vert^2   a_R(t) = -\frac{i}{2}\Gamma_{\ast}(t)\,{a}_R(t),
\end{equation}
Here, $\Gamma_{\ast}(t)$ is the time-dependent total probability for the direct ionization of the resonance.

In the presence of   $\Gamma_{\ast}(t)$,   only   Eq.~(\ref{eq:CDE_RDN2}) from the set  of   Eqs.~(\ref{eq:CDE_RDN})  must be modified accordingly
\begin{equation}
\label{eq:CDE_RDTN2}
i\dot{ {a}}_R(t)=\left(D_R(t)-\frac{i}{2} W(t)\right)\, {a}_I(t) +\left(E_R -\frac{i}{2}[\Gamma_A+\Gamma_{\ast}(t)]-\omega \right){a}_R(t) .
\end{equation}
The term $-\frac{i}{2}\Gamma_{\ast}(t)\,{a}_R(t)$ on the right-hand side of Eq.~(\ref{eq:CDE_RDTN2}) describes leakage of the population of the resonance due to direct ionization, referred to hereafter as `\emph{RD-leakage}'. The RD-leakage is time-dependent. It competes with the two alternative  mechanisms for depopulation of the resonance, the time-independent RA-leakage (due to Auger decay) and the time-dependent coupling  with the ground state (responsible for the Rabi oscillations).

\subsection{Final equations and discussion}
\label{sec:thyFIN}

Let us introduce the time-dependent vector of the amplitudes of the total wavefunction describing the population of the various involved atomic levels, the ground state  $\vert I \rangle$, the resonance $\vert R \rangle$, and the final ionic state plus Auger electron  $\vert F \varepsilon \rangle$ 
\begin{equation}
\label{eq:fin1}
\overline{A}(t)= \left(\begin{array}{c} {a}_I(t) \\ { {a}}_R(t)\\  {a}_{F_j} (t,\varepsilon_j)   \end{array} \right),
\end{equation}
and the matrix 
\begin{equation}
\label{eq:fin2}
\hat{\mathbf{H}}(t)= \left(\begin{array}{lll}E_I-\frac{i}{2}\Gamma_{ph}(t) & D^\dag_R(t)-\frac{i}{2} W^\dag(t)&0\\
D_R(t)-\frac{i}{2} W(t) ~~~ & E_R -\frac{i}{2}[\Gamma_A+\Gamma_{\ast}(t)]-\omega ~~~ &  0\\
 D_{D_j}(t)& V_{j}& E_{F_j}+\varepsilon _j-\omega  \end{array} \right),
\end{equation}
The final set of equations derived stepwise in the preceding sections now takes on the following compact form
\begin{equation}
\label{eq:fin3}
i \dot{\overline{A}}(t)= \hat{\mathbf{H}}(t)\,\overline{A}(t).
\end{equation}
The matrix $\hat{\mathbf{H}}(t)$ can be viewed as the effective Hamiltonian governing the electron dynamics in the RA process of an atom in an intense laser pulse.

Below we summarize the physical meaning of each term of the Hamiltonian matrix  (\ref{eq:fin2}). The ground state energy $E_I$ is augmented by the imaginary term $-\frac{i}{2}\Gamma_{ph}(t)$ describing the  leakage from the ground state due to its direct photoionization (GS-leakage; see Eq.~(\ref{eq:solut_D5})). Obviously, this term is time dependent because this leakage is only present during the pulse. Similarly, the total leakage from the resonance with dressed energy $E_R-\omega$  is provided by the imaginary part $-\frac{i}{2}[\Gamma_A+\Gamma_{\ast}(t)]$ on the respective diagonal element of $\hat{\mathbf{H}}(t)$. The first term, $-\frac{i}{2} \Gamma_A$, describes the usual time-independent leakage due to the Auger decay (the Auger decay rate or RA-leakage; see Eq.~(\ref{eq:tot_width})). The second term, $-\frac{i}{2}\Gamma_{\ast}(t)$, is the time-dependent leakage due to the direct photoionization of the resonance (RD-leakage; see Eq.~(\ref{eq:couplingsRRR6})).

Particularly interesting is the finding that the usual direct coupling $D_R(t)$ between the ground state and the resonance through the laser field (\cite{Rohringer08,Liu10,Sun10}; see Eq.~(\ref{eq:couplingsRexc})) is augmented by an additional a priori unexpected time-dependent term $-\frac{i}{2} W(t)$. This term appears only if the photoionization of the ground state and the Auger decay are simultaneously treated (see Eq.~(\ref{eq:solut_DR5})). We have called this additional coupling Leakage-Induced Complex coupling (LIC-coupling). This coupling is an indirect coupling between the resonance and the ground state mediated by the combined action of the Auger decay and photoionization of the ground state and can be interpreted as follows: The photoelectron emitted by the ground state is recaptured by the residual ion to produce the resonance state (as described by the term $-\frac{i}{2} W(t)=-i\pi D_{D}(t) V^\dag$   in Eq.~(\ref{eq:CDE_RDN2})) and reversely the Auger electron can be captured by the residual ion which then becomes the neutral atom in its ground state (as described by the term $-\frac{i}{2} W^\dag(t)=-i\pi V  D^\dag_{D}(t)$ in Eq.~(\ref{eq:CDE_RDN1})). These processes mediate a coupling between the resonance and the ground state as long as the pulse is on. This coupling damps the Rabi oscillations induced by the usual direct coupling $D_R(t)$ between the ground state and the resonance through the laser field.

Finally, the last row of the Hamiltonian matrix (\ref{eq:fin2}) describes how the studied specific final ionic state  $\vert F \varepsilon \rangle$  is populated by the direct photoionization of the ground state (matrix element  $ D_{D_j}(t) $; see Eq.~(\ref{eq:directME})) and coherently by the Auger decay of the resonance (matrix element$ V_{j}$; see Eq.~(\ref{eq:couplingsRexc})) all at a given kinetic energy $\varepsilon_j$ of the emitted electron. The zeros in the last column of the Hamiltonian matrix make clear that there is no direct coupling between the ground state and resonance on the one hand side and the final ionic states on the other. One can first compute the dynamics of the coupled ground state and resonance and then subsequently determine the electron spectrum (see also Eq.~(\ref{eq:CDE_RDN3})).

\section{Results and discussion}
\label{sec:res}

To exemplify the general theory of resonant Auger in intense X-ray fields we investigate the RA decay of the Ne 1s$^1$2s$^2$2p$^6$3p$^1$~$^1$P state (excitation energy is $867.12 \pm 0.03$~eV \cite{Coreno99}). For this example all input data is available or could be computed, and the measurement is feasible. The Auger decay of this resonance in the XFEL-field has recently been studied theoretically  in Ref.~\cite{Rohringer08} where the isolated resonant channel populating the final ionic states has been considered. Here, we would like to study the impact of all possible mechanisms indicated in   Fig.~\ref{fig:scheme}. We concentrate on two RA decay channels of different physical nature: the spectator RA decay into the  1s$^2$2s$^2$2p$^{4}(^1\mathrm{D})$3p~$^2$P and the participator decay into the 1s$^2$2s$^1$2p$^{6}$~$^2$S final states. The corresponding features are clearly resolved in the RA spectra as their binding energies are 55.83~eV and  48.48~eV  or, equivalently, the energies of the emitted electrons are  around   811.29~eV and 818.64~eV, respectively  \cite{Kivimaeki01,Shimizu00}.

The main difference between these two states consists in the very different ratios between the partial probabilities of the resonant and direct channels to populate them. This ratio is proportional  to   $\chi \propto \frac{D^2_RV^2_\varepsilon}{D^2_D\Gamma_A}$. States populated by the spectator decay, briefly denoted as spectator states, are states with two missing electrons in the occupied and one excited electron in the unoccupied orbitals. They are typically seen as weak satellite lines in the direct photoionization spectrum (in XPS spectrum) because their production by a single photon is forbidden in the one-electron approximation and they appear due to electron correlation and relaxation effects. On the other hand, their partial Auger rate is usually large since only occupied orbitals are involved in the Auger decay. Thus, for the spectator channel the ratio  $\chi $  is large, the Auger channel dominates over the respective direct ionization channel and the interference effects between these channels are expected to be small. In contrast, the participator states are main lines in the single photon XPS spectrum and, in addition, their partial Auger rates are small since it is the excited electron which is involved in the decay.  As a result, the ratio $\chi $ for a participator state is typically much smaller than that for the spectator state, and distinct interference patterns in the RA spectra can be expected.

\begin{table*}
\caption{Parameters for photoionization (PI) and resonant Auger (RA) decay of Ne utilized in the present calculations. The energy of the exciting radiation  is chosen to be $\omega=867.12$~eV. }
\footnotesize{
\begin{ruledtabular}
\begin{tabular}{llll}
State/Quantity& Value&Source & Method \\\hline
\multicolumn{1}{l}{Ground state 1s$^2$2s$^2$2p$^6$~$^1$S}  \\ 
Total direct PI cross section & $\sigma_D =0.024$~Mb& Ref.~\cite{BeckShirl}&Exeriment  \\\hline
\multicolumn{1}{l}{Resonance 1s$^1$2s$^2$2p$^6$3p$^1$~$^1$P}& \\  
Excitation energy& $E_R=867.12$~eV&Ref.~\cite{Coreno99}& Experiment\\ 
Total RA decay rate& $\Gamma_A=270$~meV&Ref.~\cite{Coreno99}& Experiment  \\
Oscillator strength& $f_R=0.0077$~a.u.&Present calculations&HF with relaxation\\
Total direct PI  cross section & $\sigma_\ast =0.025$~Mb&Present calculations&HF with relaxation \\
\hline
\multicolumn{1}{l}{Participator state  1s$^2$2s$^1$2p$^6$~$^2$S }  \\ 
Energy  of state & $E_F=48.48$~eV&Ref.~\cite{Kivimaeki01}& Experiment \\
Branching ratio for RA decay&$\chi^{part} \simeq 2.1$\%$^{a)}$  & Ref.~\cite{Kivimaeki01}& Experiment  \\
Partial RA decay rate& $\Gamma_A^{part}= 5.7$~meV&  & Combination of listed data \\
Partial direct PI  cross section & $\sigma_{part} =0.015$~Mb&  Ref.~\cite{BeckShirl}&Exeriment \\
\hline \multicolumn{2}{l}{Spectator state 1s$^2$2s$^2$2p$^{4}(^1\mathrm{D})$3p$^1$~$^2$P  } \\  
Energy of the state & $E_F=55.83$~eV&Refs.~\cite{Kivimaeki01,Shimizu00}& Experiment \\
Branching ratio for RA decay&$\chi^{spct}=10.6$\%$^{b)}$ &Refs.~\cite{Kivimaeki01,Shimizu00}& Experiment\\
Partial RA decay rate& $\Gamma_A^{spct}= 28.7$~meV& & Combination of listed data \\
Ratio of direct PI cross sections  & $\sigma_{spct}/\sigma_{part}$=4\%&Ref.~\cite{Oura08}& Experiment (at 900~eV) \\
Partial direct PI  cross section & $\sigma_{spct} =0.0006$~Mb&  &Combination of listed data \\
\end{tabular}
\par 
$^{a)}$ Present estimate from the Fig.~1 of Ref.~\cite{Kivimaeki01}.
\par
$^{b)}$ The intensities of the spectator transitions listed in Tab. 1 of Ref.~\cite{Kivimaeki01} are given relative to each other (i.e., they sum up to 100\%) and not relative to the total Auger rate of 270 meV what is needed here. Since 32\% of the total intensity of the RA spectrum is associated with shakeup processes \cite{Kivimaeki01}, the values listed have been  corrected accordingly.
\label{tab:param}
\end{ruledtabular}
}
\end{table*}

In order to illustrate the individual contributions of the different mechanisms to the RA spectra, the present calculations were performed not only employing the full formalism, but also employing systematically approximations which take into account only one or two mechanisms. We shall address the results of the full calculations as `exact' and the approximate ones as:
\begin{itemize}
\item `\emph{Resonant}' -- only the resonant channel is accounted for   as described in Sec.~\ref{sec:thyR};
\item  `\emph{Direct}'  -- only the direct ionization channel is taken into account as described in Sec.~\ref{sec:thyD};
\item `\emph{Interference}' --  both the direct and resonant channels and the interference between them are  taken into account as described in Sec.~\ref{sec:thyRD};
\item `\emph{Exact}' --  all terms are taken into account including the direct photoionization of the resonance as described in Sec.~\ref{sec:thySTAR} (see also Sec.~\ref{sec:thyFIN}).
\end{itemize}
The quantities utilized in the   present calculations are collected in Tab.~\ref{tab:param}. The calculations are performed for the  Gaussian-shaped  pulse  of a duration $\tau$ centered at $t_0$
\begin{equation}
\label{eq:shapeG}
g(t)=e^{-(t-t_0)^2/\tau^2}.
\end{equation}

In order to evaluate  the LIC-coupling $W(t)$, one has to include the manifold of all final ionic states (sum over index $j$ in Eq.~(\ref{eq:solut_DR5})). As is clear from the above, the individual contributions of all states to $W(t)$  are different in particular those of the spectator and participator states. Using the data in the literature, we were able to estimate only the modulus of the individual contributions of only one spectator and one participator state. In the absence of further data we assumed the same absolute values for all the important spectator and participator channels shown in Fig.~1 of Ref.~\cite{Kivimaeki01} and have further neglected the possibility that different contributions could have different signs. The resulting value of the LIC-coupling can be considered to be an upper bound for the true value. Our computations shown below illustrate that for Ne the impact of the LIC-coupling is moderate for the field intensities investigated. One can, however, anticipate that for other atoms and in particular for molecules the values and impact of the LIC-coupling can be substantial.

\subsection{Total electron yield}
\label{sec:Yield}

\begin{figure}
\includegraphics[scale=0.55]{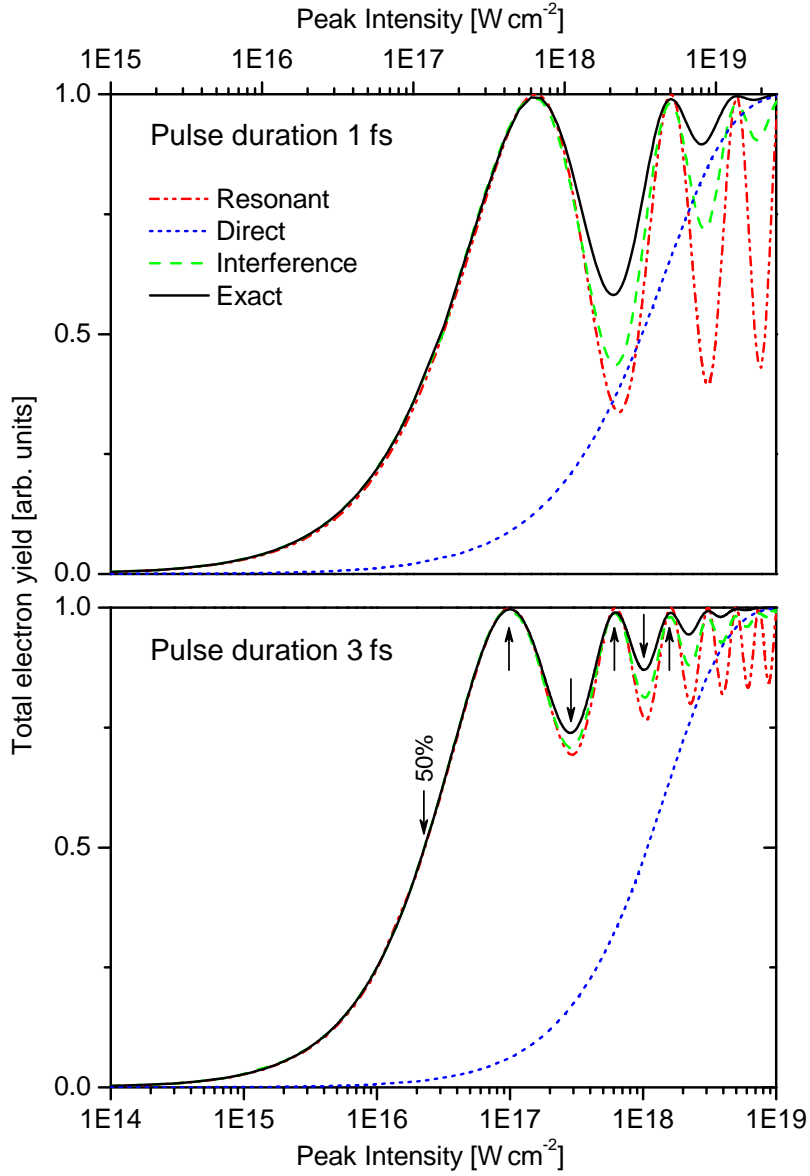}
\caption{(Color online) Total electron yield after exposure of Ne to a coherent Gaussian-shaped pulse of duration  1~fs (upper panel) and 3~fs (lower panel). Shown are `exact' results (solid line) and results of  different approximations (broken lines) discussed in the text. These include the contribution of only the resonant channel  (\emph{Resonant}; dash-dotted line), of only the direct ionization channel (\emph{Direct}; dotted line), and of both the direct and resonant channels and the interference between them (\emph{Interference}; dashed line). The upward and downward arrows in the lower panel indicate the peak intensities chosen for the calculation of the Auger spectra.}
\label{fig:yields}
\end{figure}

The total electron yield as a function of the X-ray peak intensity calculated `exactly' and compared with that obtained in the different approximations discussed above is depicted in  Fig.~\ref{fig:yields} for the two pulse durations of $\tau=1$~fs and 3~fs. The total electron yield was computed as in \cite{Rohringer08} using
\begin{equation}
\label{eq:yield}
\lim_{t\to\infty} \sum_j\int d\varepsilon_j \vert a_{F_j} (t,\varepsilon_j)\vert^2 + \lim_{t\to\infty} \sum_j\int d\varepsilon_{\ast_j} \vert{a}_{F_{\ast_j}} (t,\varepsilon_{\ast_j})\vert^2   = 1-\lim_{t\to\infty} \vert a_I (t)\vert^2.
\end{equation}
In the \emph{Resonant} approximation (dash-dotted curves), only the decay of the resonance (RA-leakage) and the Rabi oscillations between the resonance and the ground state   are competing with each other. This situation is discussed in  \cite{Rohringer08}. If the duration of the pulse is comparable or shorter than the Auger decay lifetime of the resonance ($\tau_A\sim 2.5$~fs) and its intensity is low, the atom has a finite probability of staying neutral, i.e., of remaining in its ground state, after the laser pulse is over.  As the intensity increases this probability decreases and the atom can be completely ionized with probability of 1 (see the first maximum in the total electron yield in Fig.~\ref{fig:yields}).   At certain intensities the atom manages to complete several Rabi cycles during the pulse and the ionization probability drops again. These intensities correspond to the minima in the total yield \cite{Rohringer08}, shown by the dash-dotted curves in Fig.~\ref{fig:yields}, and depend on the pulse duration (compare the upper and lower panels of this figure).

The total electron yield computed in the \emph{Direct} approximation (dotted curves) illustrates the individual contribution of the GS-leakage mechanism into the photoionization process of the atom. The X-ray pulse with energy below the 1s-threshold may ionize the 2s and 2p electrons of Ne by a single photon absorption. The total cross section for ionization of the valence shells ($\sigma_D =0.024$~Mb, Tab.~\ref{tab:param}) is much smaller than the probability for the ionization via the resonant channel. Therefore, the total electron yield saturates to 1 due to the GS-leakage at much larger intensities than in the \emph{Resonant} approximation. Because the ground state and the final ionic state posses different numbers of electrons, there are no oscillations in the total yield  (note the absence of coupling between these states in  Eqs.~(\ref{eq:CDE_DN})).

In the \emph{Interference} approximation (dashed curves in Fig.~\ref{fig:yields}), three competitive mechanisms are present in the ionization of the atom. These are the decay of the resonance (RA-leakage) and the GS-leakage discussed above, and the interference between these two channels. As discussed in the theory, the interference  is now modified from the weak field case by a complex term (LIC-coupling), as can be seen by inspecting the coupling in  Secs.~\ref{sec:thyRD} and \ref{sec:thyFIN}.  The whole coupling causes the Rabi oscillations of the population between the ground state and the resonance, and since it has now become complex, these oscillations can be modified as well and are usually suppressed. At those times when the population is in the resonant state (half-completed Rabi cycles) the RA-leakage mechanism is responsible for the ionization of the atom. On the other hand, if the population is in the ground states (completed Rabi cycles), the atom is ionized by the GS-leakage mechanism.  The competition between these two leakage mechanisms results in distinct modifications of the yield: the oscillations are now less pronounced and saturate along the trend imposed by the \emph{Direct} approximation (cf the dashed-dotted and dashed curves in  Fig.~\ref{fig:yields}). Consequently, the oscillations are suppressed from below. In addition, as can be seen in  Fig.~\ref{fig:yields}, the oscillations are also slightly suppressed from above and not all of them arrive at unity, and this weak but qualitatively interesting effect is a consequence of the complex coupling term (LIC-coupling) mentioned above.

The RD-leakage mechanism enhances the ionization of the atom (\emph{Exact} calculations, solid curves in Fig.~\ref{fig:yields}). Obviously, its impact is largest at the half-completed Rabi cycles, when the population is mainly in the resonance state, and is reduced by this leakage by ionization into other final ionic states thus suppressing somewhat the RA decay. The oscillations in the total electron yield now become even less pronounced and saturate much earlier as a function of the laser intensity (cf, dashed and solid curves). We note that depending on the pulse duration and the parameters of the system, there is a certain interval of intensities where both the GS-leakage and RD-leakage mechanisms are comparatively weak. In the present case of the longer pulse, these are the intensities up to the first minimum in the electron yield, before the atom manages to complete one Rabi cycle.

\subsection{Spectator Auger decay}
\label{sec:Spectator}

In order to illustrate the effect of the interference between the resonant and direct ionization channels we have computed the RA spectra for spectator and participator Auger decay states. The calculations were performed for the full equations and also for all systematic approximations discussed above employing a pulse duration of $\tau=3$~fs. The peak intensities chosen for the calculations correspond to the first three maxima and two minima in the total yield and also to the intensity related to the 50\% of the yield as indicated by the upward and downwards arrows in the lower panel of  Fig.~\ref{fig:yields}.

\begin{figure}
\includegraphics[scale=0.55]{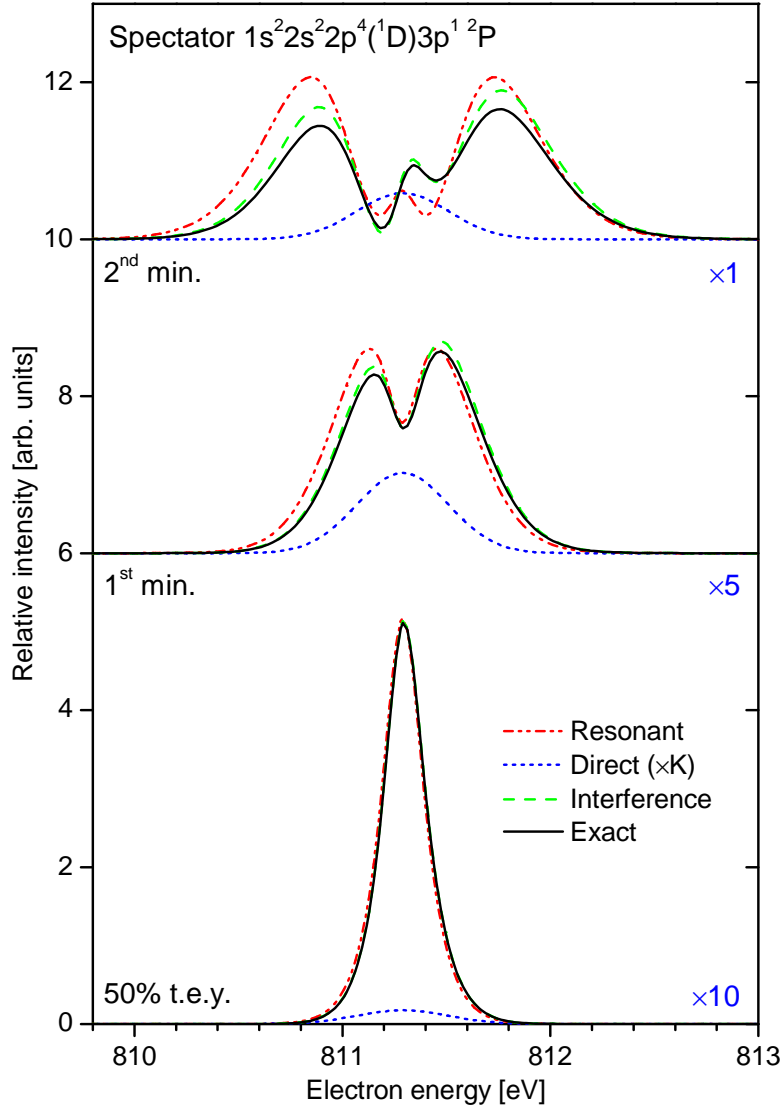}
\caption{(Color online) The spectator Auger electron spectra computed for a pulse duration of 3~fs employing the full equations (`exact') and the systematic approximations discussed in the text. The peak intensities chosen for the calculations are indicated by the downward arrows in the lower panel of  Fig.~\ref{fig:yields}. Note that the dotted curves are shown on an enhanced scale compared to the other curves indicated by the factor $\times K$  at the right hand side of the curves. }
\label{fig:spect_min}
\end{figure}

\begin{figure}
\includegraphics[scale=0.55]{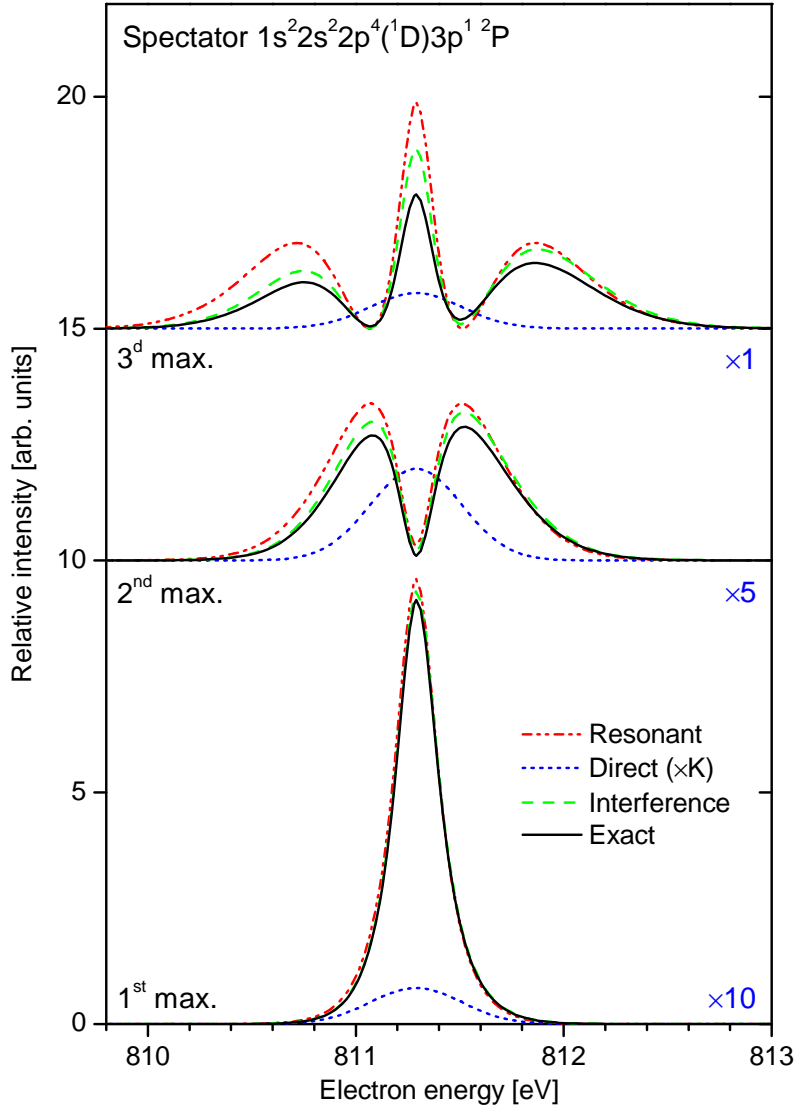}
\caption{(Color online) The spectator Auger electron spectra computed for a pulse duration of 3~fs employing the full equations (`exact') and the systematic approximations discussed in the text. The peak intensities chosen for the calculations are indicated by the upward arrows in the lower panel of  Fig.~\ref{fig:yields}. Note that the dotted curves are shown on an enhanced scale compared to the other curves indicated by the factor $\times K$  at the right hand side of the curves. }
\label{fig:spect_max}
\end{figure}

We first discuss the case of spectator RA decay. As was mentioned above, the ratio of the partial probabilities for the resonant and direct channels, $\chi \propto \frac{D^2_RV^2_\varepsilon}{D^2_D\Gamma_A}$, is large for spectator RA decay. Therefore, one intuitively expects here only small or moderate interference effects between the resonant and direct photoionization channels. The RA spectra computed for the spectator channel are depicted in Figs.~\ref{fig:spect_min} and \ref{fig:spect_max}. As was shown in Ref.~\cite{Rohringer08}, the Rabi flopping induced by a coherent X-ray pulse leads to a strong modification of the purely resonant Auger electron line profile (dash-dotted curves in the figures). With the increase of the peak intensity, the single RA electron line found at rather weak fields bifurcates resulting in local minima in the line shape. This effect was also discussed in a different context 25 years ago  \cite{Rzazwski85}.

The individual contributions of the partial direct photoionization of the ground state to the spectral line discussed are shown by the dotted curves in the figures (note that these curves are shown on an enhanced scale compared to the other curves indicated by a factor $\times K$). Since there is no coupling between the ground and the final states (see Eqs.~(\ref{eq:CDE_DN})), the line profile computed in the \emph{Direct} approximation remains unchanged with increasing laser intensity and hence displays no bifurcations. The intensity of the line in the spectrum increases rapidly as the field intensifies. Surprisingly, at peak intensities around the second minimum and the third maximum in the total electron yield (see  Fig.~\ref{fig:yields}), the individual contribution of the direct channel becomes about the same order of magnitude as that of the resonant channel as seen by comparing the dash-dotted and dotted curves in the uppermost spectra of Figs.~\ref{fig:spect_min} and \ref{fig:spect_max}. This finding can be rationalized. As seen from  Fig.~\ref{fig:yields}, the effect of the resonant channel rises fast with the intensity of the field and except of the impact of the Rabi oscillations saturates already at the first maximum of the total electron yield. The effect of the direct channel, on the other hand, becomes relevant only at higher field intensities and saturates much later. As soon as the latter is also saturated, the effect of the direct channel on the spectrum becomes comparable to that of the purely resonant channel.

The interference between the resonant and direct channels (\emph{Interference} approximation; dashed curves) results in a slight asymmetry of the RA line profiles for the spectator state. With the present parameters for the specific transition studied here, the interference is destructive on the low electron energy side, and constructive on the high energy side. Whether the interference is destructive or constructive depends, of course, on the relative signs of the corresponding amplitudes. In the present model we assumed that the resonant and the direct amplitudes have the same signs. If these signs were different, the computed spectrum would be just the mirror reflection of the line profile at the center.

Obviously, the role of the direct photoionization, and, hence, of the interference effects increases with the field strength. However, as the total electron yield saturates as a consequence of the GS-leakage mechanism, also the impact of the interference effects saturates. As the field increases further, the spectra become overall suppressed due to the depopulation of the resonance into other usually high lying final ionic states (leakage of the resonance state). The impact of this RD-leakage is visible in the resonant Auger spectra as can be seen in the figures by comparing the results of the full calculations (solid lines) with those of the interference approximation (dashed lines) not including this leakage mechanism.

\subsection{Participator Auger decay}
\label{sec:Participator}

\begin{figure}
\includegraphics[scale=0.55]{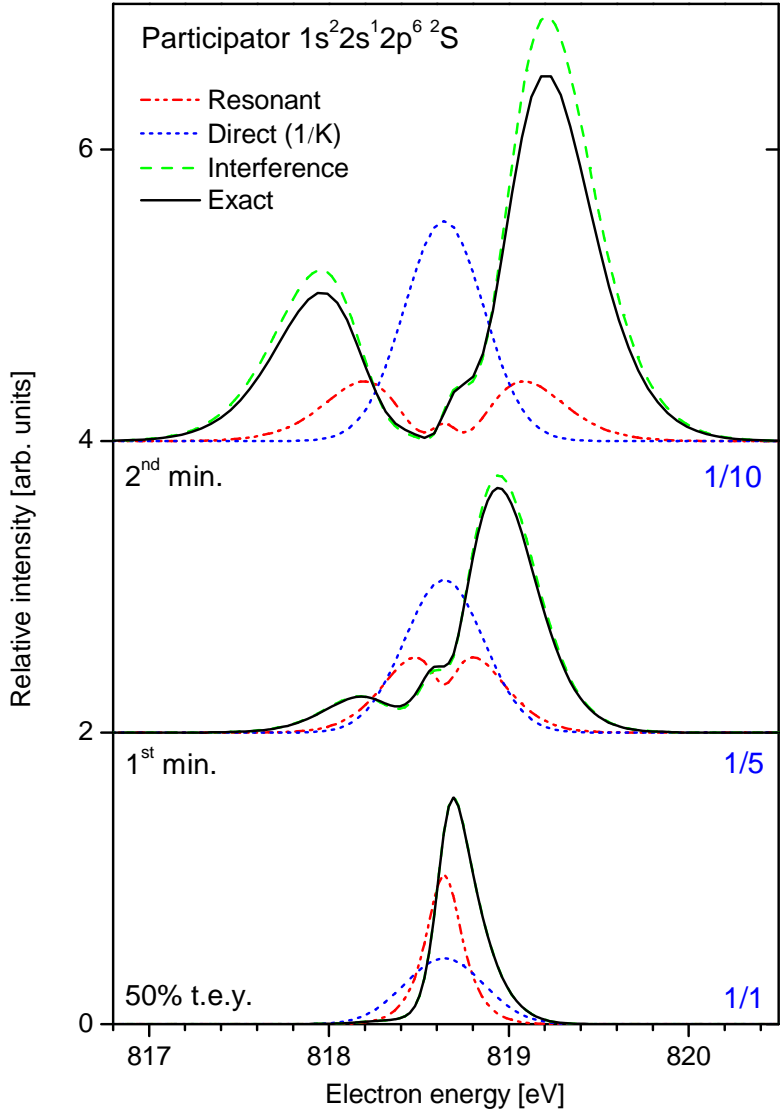}
\caption{(Color online) The participator Auger electron spectra computed for a pulse duration of 3~fs employing the full equations (`exact') and the systematic approximations discussed in the text. The peak intensities chosen for the calculations are indicated by the downward arrows in the lower panel of  Fig.~\ref{fig:yields}. Note that the dotted curves are shown on a suppressed  scale compared to the other curves indicated by the factor $ {1}/{ K}$  at the right hand side of the curves.}
\label{fig:part_min}
\end{figure}

\begin{figure}
\includegraphics[scale=0.55]{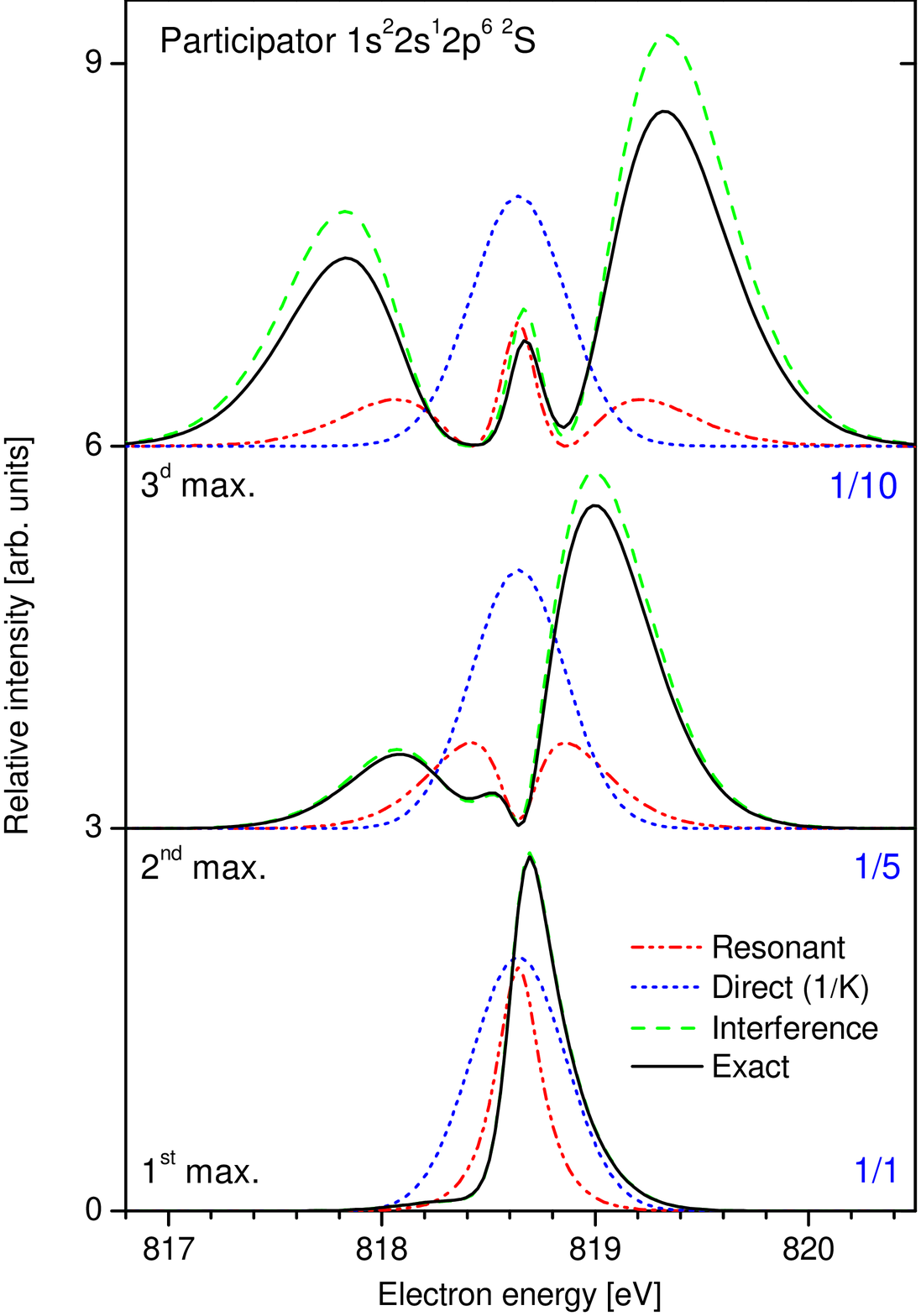}
\caption{(Color online) The participator Auger electron spectra computed for a pulse duration of 3~fs employing the full equations (`exact') and the systematic approximations discussed in the text. The peak intensities chosen for the calculations are indicated by the upward arrows in the lower panel of  Fig.~\ref{fig:yields}. Note that the dotted curves are shown on a suppressed  scale compared to the other curves indicated by the factor $ {1}/{ K}$  at the right hand side of the curves. }
\label{fig:part_max}
\end{figure}

The `exact' and approximate theoretical RA spectra of the participator channel are depicted in Figs.~\ref{fig:part_min} and \ref{fig:part_max}. The spectra computed in the \emph{Resonant} approximation show the same trends as for the spectator channel (dash-dotted curves). Because of the smaller participator Auger rate (5.7~meV compared to 28.7~meV for the spectator channel; see  Tab.~\ref{tab:param}), the participator RA electron peaks are by about 5 times weaker than those found in the spectator spectra at the same peak field intensities.

At the same time the partial direct photoionization cross section of the participator channel is 25 times larger than that of the spectator channel \cite{Oura08} ($\sigma_{part} =0.015$~Mb and  $\sigma_{spct} =0.0006$~Mb; see Tab.~\ref{tab:param}). Consequently, the individual contribution of the direct channel to the RA peaks is now seen already at relatively low field intensities (\emph{Direct} approximation; dotted curves in Figs.~\ref{fig:part_min} and \ref{fig:part_max}). The dotted curves in these figures are decreased by a factor ${1}/{ K}$ compared to the other curves in the figures. The electron spectra computed in the \emph{Resonant} and \emph{Direct} approximations are already comparable in magnitude at the intensity at which 50\% of the total ion yield is achieved for the first time in Fig.~\ref{fig:yields}. This is clearly seen when comparing the dash-dotted and dotted curves in the lowermost spectrum in Fig.~\ref{fig:part_min}. Obviously, the role of the direct channel increases greatly with the field strength as was also the case in the spectator channel. If the direct photoionization and resonant spectra were independent of each other, the direct ionization would be the dominant mechanism for the population of the participator final ionic state at the lager field intensities shown in the figures (note the factor ${1}/{ K}$).

However, interference effects are found to be substantial and the respective spectra exhibit much interesting and distinct structure. The RA spectra computed in the \emph{Interference} approximation are depicted in Figs.~\ref{fig:part_min} and \ref{fig:part_max} where they are shown as by dashed curves. By comparing these curves with the purely resonant spectra (dash-dotted curves), one immediately concludes that the interference between the resonant and direct pathways for the population of the final participator state is very important at all the studied intensities of the field. The interference patterns associated with the asymmetry of the RA spectrum are strongly pronounced. As the consequence of the present parametrization of the amplitude's signs, one observes destructive and constructive interferences on the low and high electron energy sides of the spectra, respectively (see also discussion in the preceding subsection). We also notice here that the fingerprints of the interference in the  1s$^2$2s$^1$2p$^{6}$~$^2$S participator channel might be visible even in the weak field RA spectrum measured with conventional synchrotron radiation: the slight asymmetry of the peak at binding energy of  48.48~eV in  Fig.~1 of Ref.~\cite{Kivimaeki01} is very similar to that in the lowermost spectrum in Fig. 5.

Moreover, the interference determines not only the shape of the computed RA spectra, but their integral intensities as well. Interestingly, the integral line intensity computed in the \emph{Interference} approximation at the larger intensities of the field studied (middle and uppermost spectra in  Figs.~\ref{fig:part_min} and \ref{fig:part_max}) is larger than that in the \emph{Resonant} approximation, but much smaller than that in the \emph{Direct} approximation (note factors $\frac{1}{5}$ and  $\frac{1}{10}$) . This is due to the competition between the Auger decay and GS-leakage mechanisms and persists also in the `exact' calculations where the leakage from the resonance is taken into account (solid curves). Finally, we note that the impact of the latter losses to higher ionic states (RD-leakage mechanism) on the decay spectra is to suppress their intensity without substantially changing their line shape.

\section{Conclusions}
\label{sec:conc}

The theory of resonant Auger decay of an atom exposed to a coherent monochromatic X-ray pulse of strong intensity is presented. The theory incorporates the resonant and direct photoionization pathways from the ground electronic state to the final ionic states, as well as the ionization losses of the resonance state. Interesting interference effects appear in the electron spectra due to the presence of the direct ionization pathway and the excitation and decay of the resonance which populate the final ionic state coherently. Due to the leakage of the ground state, the coupling between the ground electronic state and the `dressed' resonance state is modified giving rise to an additional complex coupling term which we call the Leakage-Induced Complex coupling (LIC-coupling). All together, the effective Hamiltonian governing the production of ions and electrons is non-hermitian where all its terms, diagonal and off-diagonal, are complex.

The theory predicts a strong competition of several mechanisms of ionization of the atom. While the pulse proceeds, the non-hermitian coupling induces (damped) Rabi oscillations of the population between the ground state and the resonance. While the population is in the ground state, in particular when a Rabi cycle is completed, it is subject to leakage into ionic states via direct photoionization (GS-leakage). In the time intervals when the population is in the resonant state, in particular at half-completed Rabi cycles, it can decay by resonant Auger decay (RA-leakage), but it can also be transferred into highly-excited ionic states via direct photoionization of the resonance itself (RD-leakage).

The developed formalism is applied to study the excitation and Auger decay of the  1s$^1$2s$^2$2p$^6$3p$^1$~$^1$P resonance of a Ne atom in the intense field of an X-ray pulse with a central frequency of  867.12~eV. It is shown that at relatively weak intensities of the field, the resonant Auger decay is the dominant mechanism of ionization of Ne, but as soon as this channel of ionization saturates with the increase of the field intensity, the GS-leakage as well as the RD-leakage mechanisms start to be competitive. At larger intensities, when the impact of the GS-leakage and RD-leakage saturates as well, all these three different mechanisms of production of ionized Ne become of comparable magnitude.

The interference effects are of different importance to spectator and participator final states of the Auger decay. To demonstrate the interference effects between the resonant and direct photoionization pathways, we have studied the electron spectra for the two specific cases of the spectator 1s$^2$2s$^2$2p$^{4}(^1\mathrm{D})$3p~$^2$P and the participator 1s$^2$2s$^1$2p$^{6}$~$^2$S final Auger states. The interference effects increase with the field strength until they saturate when the total electron yield saturates due to the GS-leakage mechanism. For the final state produced by the spectator decay, the decay by resonant Auger is the dominant ionization mechanism, and the interference results in a slight asymmetry in the line profile of the spectrum. In contrast, the direct photoionization is the dominant mechanism at large field intensities in the case of participator final states. Nevertheless, a distinct interference pattern appears in the electron spectrum which remains highly structured in spite of the strong direct photoionization at high field intensities. This finding makes the study of resonant Auger spectra in strong fields worth while.

We would like to conclude with the following remark. In molecules, the discussed competition of different leakage mechanisms and the interference effects will be much more intricate, due to the presence of the nuclear dynamics. In particular, we expect in molecules also the appearance of strong non-adiabatic effects induced by the field \cite{dices1,dices2}. Here, we suggest to follow the emerging competition and interference effects in the time domain which will hopefully be possible with the advances in the development of pump-probe techniques utilizing femtosecond and sub-femtosecond pulses.

\begin{acknowledgments}
A. I. Kuleff, S. D. Stoychev and Y.-C. Chiang are gratefully acknowledged for many useful discussions. Financial support by the DFG is acknowledged.
\end{acknowledgments}

\appendix
\section{Local approximation}
\label{sec:app}

In this appendix we discuss the local approximation which leads to $\Gamma_{ph}(t)$ in Eq.~(\ref{eq:solut_D5}) as an example. The starting point is the set of equations (\ref{eq:CDE_D}). The straightforward integration of this set is impossible owing to the continuous set of Eqs.~(\ref{eq:CDE_D2}) describing the final ionic state together with the photoelectron. In order to evaluate the integral in  Eq.~(\ref{eq:CDE_D1}), one would have to solve the whole set of Eqs.~(\ref{eq:CDE_D2}) simultaneously for all electron energies which is a formidable task.  The problem becomes tractable by implying the local approximation  \cite{Cederbaum81,Domcke91} to integrate Eqs.~(\ref{eq:solut_D2}) and (\ref{eq:solut_D3}). Due to the non-locality of the kernel (\ref{eq:solut_D3}), the amplitude $a_I(t)$ is needed at all times smaller than $t$, i.e., the kernel is a so-called `memory kernel' \cite{Domcke91}. The local approximation consists in eliminating this `memory' by localization of the kernel (\ref{eq:solut_D3}). In order to proceed, we make the same assumptions as in Ref.~\cite{Pahl99ZPD} where the local approximation is derived in detail.

At first, we assume that the integral function  $D^\dag_D(t) D_D(t^\prime) $ in Eq.~(\ref{eq:solut_D3}) varies negligibly as a function of energy in the symmetric energy interval of  ($\varepsilon_0-\epsilon,\varepsilon_0+\epsilon$)  around the position of the center of the  spectral line of the emitted electron $\varepsilon _0 =E_I+\omega-E_F$, and vanishes otherwise (the choice of the interval length,  parameter $\epsilon$, will be justified below). The approximation is good as long as $\varepsilon_0$ is large, i.e., the ejected electron has a large energy and the ionization process is far from the ionization threshold. In this approximation we obtain
\begin{multline}
\label{app1}
 K(t-t^\prime)= \int_0^\infty  D^\dag_D(t)  \, D_D(t^\prime)   e^{-i(E_F+\varepsilon_0-\omega )(t-t^\prime)}    e^{-i(\varepsilon-\varepsilon_0 )(t-t^\prime)}  d\varepsilon =\\
 D^\dag_D(t)  \, D_D(t^\prime)   \,  e^{-i(E_F+\varepsilon_0-\omega )(t-t^\prime)}  \int_{-\epsilon}^\epsilon  e^{-ix(t-t^\prime)}  dx=\\
 \frac{2\sin\left[\epsilon(t-t^\prime)\right]}{t-t^\prime}D^\dag_D(t)   D_D(t^\prime)  e^{-i(E_F+\varepsilon_0-\omega )(t-t^\prime)} .
\end{multline}
Next, we assume that the time-dependence of the amplitude $a_I(t)$ can be factorized in two parts of different physical origins  
\begin{equation}
\label{app2}
a_I(t)= e^{-iE_It}a^{(0)}_I(t).
\end{equation}
The phase factor $e^{-iE_It}$ describes the time-dependence of the amplitude owing to the energy of the ground state.  The rest of the time-dependence of the amplitude $a_I(t)$ is associated with  the impact of the laser pulse, and is deposited in the function  $a^{(0)}_I(t)$.

The time integral in Eq.~(\ref{eq:solut_D2}) can now be simplified 
\begin{multline}
\label{app3}
-i \int_{-\infty}^t  K(t-t^\prime) a_I(t^\prime)  dt^\prime = -i  \int_{0}^\infty K(\tau)  a_I(t-\tau)  d\tau=\\
-i D^\dag_D(t) \int_{0}^\infty   \frac{2\sin\left[\epsilon\tau\right]}{\tau}   D_D(t-\tau)  e^{-i(E_F+\varepsilon_0-\omega -E_I)\tau} e^{-iE_It}a^{(0)}_I(t-\tau)    d\tau=\\
-i D^\dag_D(t)  e^{-iE_It} \int_{0}^\infty   \frac{2\sin\left[\epsilon\tau\right]}{\tau}   D_D(t-\tau)a^{(0)}_I(t-\tau)    d\tau,
\end{multline}
where in the last step we used the definition of the center of the electron spectral line $E_F+\varepsilon_0-\omega -E_I=0$. The main difference in  further integrating Eq.~(\ref{app3})   from the procedure described in Ref.~\cite{Pahl99ZPD} consists in the time dependence of the $D_D(t)$-function. To proceed we note that the timescales at which the functions in the integrand undergo substantial variations depend on the pulse duration and on the yet to be determined parameter $\epsilon$.  The function $D_D(t-\tau)a^{(0)}_I(t-\tau)$ in Eq.~(\ref{app3}) varies  on the timescale of the pulse duration (usually a few femtoseconds). The relevant timescale for substantial changes of the other appearing  function, $ \frac{2\sin\left[\epsilon\tau\right]}{\tau}$, is determined by the parameter $\epsilon$  and is of the order of  $\frac{2\pi}{\epsilon}$. In order to be able to perform  the integration of Eq.~(\ref{app3}) over time analytically,  one has to make the latter timescale much shorter than the duration of the pulse. For this purpose, the parameter $\epsilon$  has to be taken as large as possible.

The upper limit for the parameter $\epsilon$ is naturally determined by the ionization threshold ($\epsilon \le \varepsilon_0$).  In the case of core-excitation, the electron energy  $\varepsilon_0$ is large (in the present case more than 800~eV). By taking the parameter $\epsilon=\varepsilon_0$ (as in Ref.~\cite{Pahl99ZPD}), the timescale for  substantial variations of the function $\frac{2\sin\left[\epsilon\tau\right]}{\tau}$ can be made much shorter than the duration of the pulse. For the present case, this timescale is in the sub-femtoseconds domain  ($\frac{2\pi}{\varepsilon_0}\approx 0.005$~fs).  Finally, we note that the function $\frac{2\sin\left[\epsilon\tau\right]}{\tau}$ vanishes rapidly as $\tau$ increases. As a result, the main contribution in the integral (\ref{app3}) comes from the times around $\tau \approx 0$. The analytical integration of Eq.~(\ref{app3}) is now straightforward and we obtain 
\begin{equation}
\label{app4}
-i D^\dag_D(t) D_D(t)  e^{-iE_It}a^{(0)}_I(t) \int_{0}^\infty   \frac{2\sin\left[\epsilon\tau\right]}{\tau}      d\tau =-i \pi \vert D_D(t) \vert^2 a_I(t) ,
\end{equation}
and hence Eq.~(\ref{eq:solut_D4}) as the final result.

%\bibliographystyle{apsrev}
%\bibliography{correlat,molec,mybooks,N2,no,other,ours,ResAuger,karl,rydbergs,dissertation,comol,cederbaum,xfel}

\end{document}